\documentclass[11pt]{article}

\usepackage{amssymb,amsmath}

\usepackage{amsmath}
\usepackage{amssymb}
\usepackage{amsthm}
\usepackage{url}

\newcommand\x{{\bf x}}

\newcommand\z{{\bf z}}

\newcommand\zero{{\bf 0}}

\newtheorem{theorem}{Theorem}[section]

\newtheorem{algorithm}[theorem]{Algorithm}

\theoremstyle{definition}

\begin{document}

\title{
Quality Up in Polynomial Homotopy Continuation by
Multithreaded Path Tracking\thanks{This
material is based upon work supported by the National Science Foundation
under Grant No.\ 0713018 and Grant No.\ 1115777.}}
\author{Jan Verschelde and Genady Yoffe \\
Department of Mathematics, Statistics, and Computer Science \\
University of Illinois at Chicago \\
851 South Morgan (M/C 249) \\
Chicago, IL 60607-7045, USA \\
\texttt{jan@math.uic.edu, gyoffe2@uic.edu} \\
\texttt{\url{www.math.uic.edu/~jan} ~~~~~~~~~~~~~~~~~~~~~ }}

\date{2 September 2011}

\maketitle

\begin{abstract}
Speedup measures how much faster we can solve the same problem
using many cores.  If we can afford to keep the execution time fixed,
then quality up measures how much better the solution will be computed
using many cores.  In this paper we describe our multithreaded 
implementation to track one solution path defined by a polynomial homotopy.
Limiting quality to accuracy and confusing accuracy with precision,
we strive to offset the cost of multiprecision arithmetic running
multithreaded code on many cores.
\end{abstract}

\section{Introduction}

Solving polynomial systems by homotopy continuation proceeds
in two stages: we first define a family of systems (the homotopy)
and then we track the solution paths defined by the homotopy.
Tracking all paths is a pleasingly parallel computation.
The problem we consider in this paper is to track one solution path.
While tracking only one solution path could occur for huge problems
(for which it is no longer feasible to compute all solutions), or
the need to track one difficult solution path for which multiprecision
arithmetic is required often arises for larger systems.

On a multicore workstation, we experimentally determined in~\cite{VY10}
thresholds on the dimension of the problem to achieve a good speedup for
the components of Newton method, using the quad double software library
QD-2.3.9~\cite{HLB00}.  While polynomial evaluation often dominates the
computational cost, it pays off to run also multithreaded versions of
the Gaussian elimination stage of Newton's method.
In this paper we describe our multithreaded path tracker.

The idea to use floating-point arithmetic as implemented in QD-2.3.9
to increase the working precision dates back to~\cite{Dek71}.
In~\cite{Rum10}, this idea is described as an error-free transformations 
because with double doubles we can calculate the error of a
floating-point operation.  From a computational complexity point
of view, double doubles are attractive because the cost overhead
is similar to working with complex arithmetic.
For multithreading -- using concurrent tasks accessing shared
memory -- double doubles are very attractive because blocking
memory allocations and deallocations do not occur.

As in~\cite{VY10} we continue the development of double double and
quad double arithmetic in our path trackers and experimentally
determine thresholds on the dimensions and degrees to achieve
a good speedup.  In the next section we relate our calculations
with the less commonly used notion of quality up~\cite{Akl04}.
As double floating-point arithmetic has become the norm,
we can ask how much faster our hardware should become for
double double arithmetic to become our default precision?
We illustrate the computation of quality up factors in the next section.

In~\cite{VY10} we experienced that for homotopy continuation methods,
polynomial evaluation is the dominating cost factor, exceeding the
cost of the linear system solving as required in Newton's method
-- although it still pays off to multithread Gaussian elimination.
For Newton's method we not only need to evaluate polynomials but
also all derivatives with respect to all unknowns are needed in
the Jacobian matrix.  Using ideas from algorithmic differentiation~\cite{GW08}
we have been able to reduce the dominating cost factor.

Another application area for the techniques of this paper
is the deflation of isolated singularities~\cite{LVZ06}.
The deflation method accurately locates singular solutions
at the expense of adding higher derivatives to the original system,
essentially doubling the dimension in every stage.
Any implementation of this deflation will benefit from
increased precision and efficient evaluation of polynomials and
all their derivatives.  In this setting the granularity of the
parallelism must be fine and the use of multithreading is needed.

\noindent {\bf Acknowledgements.}  The ideas for the quality up section
below were developed while preparing for an invited talk of the second
author at the workshop on Hybrid Methodologies for Symbolic-Numeric
Computation, held at MSRI from 17 to 19 November 2010.  The second
author is grateful to the organizers of this MSRI workshop for
their invitation.

\section{Speedup and Quality Up}

When using multiple cores, we commonly ask how much faster
we can solve a problem when using $p$ cores.
Denoting $T_p$ the time on $p$ cores, then the speedup is
defined as $T_1/T_p$, i.e.: the time on 1 core divided by
the time on $p$ cores.  In the optimal case, the speedup
will converge to~$p$: with $p$ cores we can solve the same
problem $p$ times faster.

In addition to speedup, we like to know how much better 
we can solve the problem when using $p$ cores?
Our notion of quality up as an analogue to speedup is inspired
by Selim Akl's paper~\cite{Akl04}.
We define quality as the number of correct decimal places in
the computed solution.  Denoting $Q_p$ as the quality obtained
using $p$ cores, we define quality up as $Q_p/Q_1$, keeping the
time fixed.  As with speedup, we could also hope for a quality
up factor of~$p$ in the optimal case.

Because the number of correct decimal places in a numerical solution
depends on the sensitivity of the solution to perturbations in the input
and is (bounded by condition numbers, we assume our problems are
well-conditioned deliberately confusing working precision with accuracy. 
Taking a narrow view on quality, we define
\begin{displaymath}
  {\rm quality~up} = \frac{Q_p}{Q_1}
   = \frac{\mbox{\rm \# decimal places with $p$ cores}}
          {\mbox{\rm \# decimal places with 1 core}}.
\end{displaymath}
Often multiprecision arithmetic is necessary to obtain meaningful answers
and then we want to know how many cores we need to compensate for the 
overhead caused by software driven arithmetic.
Using the quad double software library QD-2.3.9~\cite{HLB00},
we experimentally determined in~\cite{VY10} that the computational overhead
of using double double arithmetic over hardware double arithmetic
on solving linear systems with LU factorization averaged around eight.
This experimental factor of eight is about the same overhead factor of
using complex arithmetic compared to real arithmetic.
The number eight also equals
the number of cores on our Mac OS X 3.2 Ghz Intel Xeon workstation.

To estimate the quality up factors, we assume an optimal (or constant)
speedup.  Moreover, we assume that the ratio $Q_p/Q_1$ is linear in~$p$
so we can apply linear extrapolation.
To illustrate the estimation of the quality up factor, consider the
refinement of the 1,747 generating cyclic 10-roots.
The cyclic 10-roots problem belongs to a well known family of benchmark
polynomial systems, see for instance~\cite{DKK03}, \cite{LLT08}, or~\cite{Ver99}.
To compare quality up, we compare the {\tt 4.818} seconds of real time
with one core using double double complex arithmetic to the {\tt 8.076}
seconds of real time using quad double arithmetic using 8 cores.
With 8 cores we double the accuracy in less than double the time.
As this refinement is a pleasingly parallel calculation,
the assumption that the speedup is optimal is natural.
The concept of quality up requires a constant time, so we ask:
how many cores do we need to reduce
the calculation with quad doubles to {\tt 4.818s}?

\begin{displaymath}
  \frac{8.076}{4.818} \times 8 = 13.410
  \Rightarrow 14 ~{\rm cores}
\end{displaymath}

Denoting $y(p) = Q_p/Q_1$ and assuming $y(p)$ is linear in~$p$,
we have $y(1) = 1$ and $y(14) = 2$, so we interpolate:
\begin{displaymath}
   y(p) - y(1) = \frac{y(14) - y(1)}{14 - 1}(p - 1).
\end{displaymath}
and the quality up factor is
${\displaystyle y(8) = 1 + \frac{7}{13} \approx 1.538}$.
The interpretation for the factor 1.538 is as follows:
in keeping the total time fixed, we can increase the working
precision with about 50\% using 8 cores.

\section{Multithreaded Path Tracking}

Given a homotopy~$h(\x,t) = \zero$ and a start solution at~$t = 0$,
the path tracker returns the solution at the path at~$t = 1$. 
A path tracker has three ingredients: a predictor for the next
value of~$t$ and the extrapolated corresponding values for~$\x$;
Newton's method as a corrector, keeping the predicted value for~$t$ fixed;
and a step size control algorithm.  

Algorithm~\ref{algmtpath} provides
pseudo code for the multithreaded version of a path tracker.
Every thread executes the same code.
Variables that start with {\tt my\_} are local to the thread.
The first thread manages the flags used for synchronization. 
Because threads are created once and remain allocated to the
path tracking process till the end, idle threads are not released
to the operating system, but run a busy waiting loop.

\begin{algorithm}[Multithreaded Path Tracking] \label{algmtpath} {\rm
~~
\begin{center}
\begin{tabular}{lcr}
Input: $h(\x,t) = \zero$, $\z$. & & {\em homotopy and start solution} \\
Output: $\z$: $h(\z,1) = \zero$ or fail.
  & & {\em solution at end of path or failure} \\
\\
stop := false; & & {\em initializations} \\
$\lambda$ := initial step size; & & {\em all other variables are set to 0} \\
while ($t<1$ and not stop) do \\
\hspace{3mm}  while (corr\_Ind $<$ {\tt my\_corr\_Ind} ) wait;
  & & {\em wait till previous post correction} \\
\hspace{3mm} if ({\tt my\_ID} = 1) then 
  & & {\em prediction done by thread 1}  \\
\hspace{6mm} predict$(\z,t,\lambda)$;  
  & & {\em new $\z$ and t} \\
\hspace{6mm} pred\_Ind := pred\_Ind+1;
  & & {\em signal that prediction done} \\
\hspace{3mm} end if; \\
\hspace{3mm} while (pred\_Ind $<$ corr\_Ind+1) wait;
  & & {\em wait till prediction done} \\
\hspace{3mm} Newton({\tt my\_ID}, $h$, $\z$, $\epsilon$, Max\_It, success);
  & & {\em run multithreaded Newton} \\
\hspace{3mm} Newton\_Ind[{\tt my\_ID}]
  := Newton\_Ind[{\tt my\_ID}] + 1; & & {\em thread {\tt my\_ID} is done} \\
\hspace{3mm} while ($\exists$ ID: Newton\_Ind[ID] $<$ corr\_Ind+1) wait;
  & & {\em wait till correction terminates} \\  
\hspace{3mm} if ({\tt my\_ID} = 1) then 
  & & {\em step size control by thread 1}  \\
\hspace{6mm} step\_size\_control($\lambda$, success);
  & &  {\em adjust step size} \\
\hspace{6mm} step\_back($\z$, t, success); 
  & & {\em step back if no success} \\
\hspace{6mm} stop := stop\_criterion($\lambda$,corr\_Ind);
  & & {\em $\lambda$ too small or corr\_Ind too large} \\ 
\hspace{6mm} corr\_Ind := corr\_Ind + 1;
  & & {\em step size control is done} \\
\hspace{3mm} end if; \\
\hspace{3mm} my\_corr\_Ind := my\_corr\_Ind + 1;
  & & {\em continue to next step in while} \\
end while; \\
fail := not stop.
  & & {\em failure if stopped with $t < 1$}
\end{tabular}
\end{center}
}
\end{algorithm}

Prediction and step size control are relatively inexpensive
operations and are performed entirely by the first thread.
Newton's method is computationally more involved and is
executed in a multithreaded fashion.

\section{Multithreaded Newton Method}

In this section we focus on our multithreaded version of
Newton's method using multithreaded polynomial evaluation
and linear system solving described in~\cite{VY10}.
Following the same notational conventions as in
Algorithm~\ref{algmtpath}, 
pseudo code is described in Algorithm~\ref{algmtnewt} below.

\newpage

\begin{algorithm}[Multithreaded Newton's Method] \label{algmtnewt} {\rm
~~
\begin{center}
\begin{tabular}{lcr}
Input: $h(\x,t) = \zero$, $\z$; & & {\em homotopy and initial solution} \\
\hspace{11mm} $\epsilon$, Max\_It. 
  & & {\em tolerance and maximal \#iterations} \\
Output: $\z$: $||h(\z,t)|| < \epsilon$ or fail.
  & & {\em corrected solution or failure} \\
\\
$i := 0$; & & {\em count \#iterations} \\
$||h(\z,t) || := 1$; & & {\em initialize residual} \\
while ($||h(\z,t)|| > \epsilon$) and ($i <$~ Max\_It) do \\
\hspace{3mm} $V$ := Monomial\_Evaluation({\tt my\_ID},$h$,$\z$);
  & & {\em multithreaded monomial evaluation} \\
\hspace{3mm} Status\_MonVal[{\tt my\_ID}] := 1;
  & & {\em thread done with monomial evaluation} \\
\hspace{3mm} if ({\tt my\_ID} = 1) then 
  & & {\em flag adjustments for next stage} \\
\hspace{6mm} while ($\exists$ ID: Status\_MonVal[ID] = 0) wait; 
  & & {\em thread 1 waits} \\
\hspace{6mm} for all ID do Status\_MonVal[ID] := 0; 
  & & {\em flags reset for next stage} \\
\hspace{6mm} Mon\_Ind := Mon\_Ind + 1; & & {\em update monomial counter} \\
\hspace{3mm} end if; \\
\hspace{3mm} while (Mon\_Ind $<$ my\_Iter+1) wait; 
  & & {\em wait till all monomials are evaluated} \\
\hspace{3mm} $Y$ := Coefficient\_Product({\tt my\_ID},$V$,$h$);
  & & {\em multiply monomials with coefficients} \\
\hspace{3mm} Status\_Coeff[{\tt my\_ID}] := 1;
  & & {\em thread done with coefficient product} \\
\hspace{3mm} if ({\tt my\_ID} = 1) then
  & & {\em flag adjustments for next stage} \\
\hspace{6mm} while ($\exists$ ID: Status\_Coeff[ID] = 0) wait;
  & & {\em thread 1 waits} \\
\hspace{6mm} for all ID do Status\_Coeff[ID] := 0;
  & & {\em flags reset for next stage} \\
\hspace{6mm} $||h(\z,t)|| := $ Residual($Y$);
  & & {\em calculate residual} \\
\hspace{6mm} Coeff\_Ind := Coeff\_Ind + 1;
  & & {\em update coefficient counter} \\
\hspace{3mm} end if; \\
%
\hspace{3mm} while (Coeff\_Ind $<$ my\_Iter+1) wait;
  & & {\em wait till all polynomials are evaluated} \\
\hspace{3mm} $Ab$ := GE({\tt my\_ID},my\_Iter,$Y$,pivots);
  & & {\em row reduction on Jacobi matrix} \\
\hspace{3mm} $m := (n-1)$(my\_Iter+1); 
  & & {\em used for synchronization} \\
\hspace{3mm} while ($\exists$ ID: pivots[ID] $< m$) wait;
  & & {\em wait for row reduction to finish} \\
\hspace{3mm} Back\_Subs({\tt my\_Id},my\_Iter,$Ab$,$\Delta \z$,BS\_Ind);
  & & {\em multithreaded back substitution} \\
\hspace{3mm} while (BS\_Ind $<$ my\_Iter+1) wait;
  & & {\em wait till back substitution done} \\
\hspace{3mm} if ({\tt my\_ID} = 1) then \\
\hspace{6mm} $\z := \z + \Delta \z$; $i := i + 1$;
  & & {\em update solution} \\
\hspace{6mm} $\z$\_Ind := $\z$\_Ind + 1;
  & & {\em counter to update~$\z$} \\
\hspace{3mm} end if; \\
\hspace{3mm} while ($\z$\_Ind $<$ my\_Iter+1) wait;
  & & {\em wait till solution is updated} \\
\hspace{3mm} my\_Iter := my\_Iter + 1; \\
end while; \\
fail := $||h(\z,t)|| \geq \epsilon$.
\end{tabular}
\end{center}
}
\end{algorithm}

The array~$Y$ contains the evaluated polynomials of the
polynomial system as defined by the homotopy~$h(\x,t) = \zero$
along with all partial derivatives as needed in the Jacobian matrix.
The evaluation of the all polynomials as described in~\cite{VY10}
occurs in two stages: first the values of all monomials 
are stored in~$V$ and then we multiply with the coefficients
to obtain~$Y$.  The partitioning of the work load between the
threads is such that no synchronization within the procedures
Monomial\_Evaluation and Coefficient\_Product is needed.

The array~$Y$ contains then all the information needed to set up
the linear system $A \x = {\bf b}$.  The row reduction with pivoting
is performed on the augmented matrix $[A~{\bf b}]$, denoted in the
algorithm by~$Ab$.   For numerical stability, we apply pivoting in
the routine GE of Algorithm~\ref{algmtnewt}.  The pivoting implies
that within the procedure GE synchronization is needed.
For synchronization in GE, we follow the same protocol:
the first thread selects the pivot element (the largest number
in the current column).  After the selection of the pivot row,
all threads can update their preassigned part of the matrix.
The assignment of rows relates the row number ot the identification
number of the thread.  The selection of the pivot row must wait
till all threads have finished modifying their rows.

The output of the procedure GE is passed 
to the back substitution procedure Back\_Subs.
As the back sustitution solves a triangular system,
inside the routine synchronization is necessary for correct results.

To project the speedup for path tracking a system,
we generate a system of 40 variables
with a common support of 200 monomials.
Every monomial has degree 40 on average with 80 as largest degree.
We simulated 1,000 Newton iterations
and results are reported in Table~\ref{tabsimulation}.

\begin{table}[hbt]
\begin{center}
\begin{tabular}{c||r|r|r|r||c}
\multicolumn{6}{c}{\tt 40-by-40 system, 1000 times} \\
\#threads & \multicolumn{1}{c|}{Pol.Ev.} &
\multicolumn{1}{c|}{Gauss.El.} & \multicolumn{1}{c|}{Back Subs.} &
\multicolumn{1}{c||}{Total}&  speedup \\ \hline
1 & {\tt 35.732s} & {\tt 4.849s} & {\tt 0.197s} & 40.778s & 1 \\
2 & {\tt 17.932s} & {\tt 3.113s} & {\tt 0.100s} & 21.145s& 1.928 \\
4 & {\tt 9.248s} & {\tt 1.824s} & {\tt 0.062s} & 11.134s& 3.662 \\
8 & {\tt 4.775s} & {\tt 1.349s} & {\tt 0.053s} & 6.177s& 6.602 \\
\end{tabular}
\caption{Elapsed wall clock time for increasing number of threads for
polynomial evaluation, Gaussian elimination and back substitution.
The speedup is calculated for the total time.  }
\label{tabsimulation}
\end{center}
\end{table}

For the generated problem the polynomial evaluation
dominates the total cost.  In Table~\ref{tabsimulation} we see
that once we reduced the cost of polynomial evaluation using 8 cores,
the wall clock time becomes less than the total time spent on Gaussion
elimination with one core.  While multicore row reduction has a less
favorable speedup compared to polynomial evaluation, we see that
the multithreaded version is beneficial for the total speedup.

\section{Effect of a Quadratic Predictor}

Our multithreaded implementation achieves the better speedups
the bigger is the ratio of dimension 
of the system to the number of engaged cores.
Thus we work with larger dimensions. Secant predictor, which is 
merely efficient for systems of smaller dimensions becomes extremely 
unefficient for larger dimensions.
We need to come up with a predictor, which would better approximate 
local intricate behaviour of a curve in multidumensional space. 
A suitable option for this proved to be
the following quadratic predictor: 
\begin{itemize}
\item Tracking a path, we keep approximate solutions
      $x^*$, $x_{prev}^*$, and $x_{prev1}^*$ of intermediate systems,
      corresponding to the three most recent values of the homotopy parameter
      $t_{prev1} < t_{prev} < t$ respectively.
\item For each index $i$, the coordinate $x^*[i]$ of the new initial 
      guess $x^*$ for the solution on the path of the intermediate
      system associated with the value of the homotopy parameter
      ($t$ + current step size),  is computed independently
      of other coordinates as following:

\begin{enumerate}
\item We interpolate points $(t_{prev1}, x_{prev1}^*[i])$, 
      $(t_{prev}, x_{prev}^*[i])$, and $(t, x^*[i])$
      by a parabola.

\item The new $x^*[i]$ is then the value of this parabola
      at the point ($t$ + current step size).
\end{enumerate}
\end{itemize}

Since each coordinate of such guess for a new intermidiate system 
is computed independently of all the others, and all what it requires is 
computing just one value of interpolating three points parabola, 
which is done by a finite fixed number of algebraic operations,
the complexity 
of such predictor depends linearly on the dimension of the system.  
Thus the portion of quadratic predictor computation in the entire 
path tracker computation is negligable.
There is no reason to multitask quadratic predictor therefore, 
despite apparently it could be effectevely done with a minimal effort.
On the other hand, despite its very low computational time cost, 
use of the described above quadratic 
predictor instead of the secant predictor
brings dramatic gain in the number of needed corrections to track a path.
In our experiments we tracked on 8 cores a solution path for a system of 
dimension 20, with 20 monomials in each polynomial,
with each monomial of maximal degree 2
using both predictors. When using the secant predictor, it required
113623 succesful corrections , with a running time 26m53.008s,
minimal step size 6.10352e-07, and average step size 9.0404e-06,
and when using the quadractic predictor, it required 572 succesful 
corrections , with a running time 8.863s,         
minimal step size 0.00016, and average step size 0.00019.
In paricular the running time, when using quadratic predictor was about 
180 less than when using the secant predictor.
In all our other numerious experiments with systems of big enough various 
dimensions
the gain of using the quadratic predictor kept to be of the same oreder.
For a system of dimension 40 a run on 8 cores with a use of the quadratic 
predictor may take several minutes
while a run with a use of the secant predictor may take several days. 

The quadratic predictor provides very suitable balance between its 
low computational complexity and reduction in number of corrections
it brings, thus ensuring a considerable, and probably one of the 
best possible, gain in absolute running time when tracking a path.
The tables below show timings that illustrate the beneficial
effect of using a quadratic predictor.
On systems of the same dimension and degrees,
Table~\ref{tabsecant} shows experimental results
of runs with a secant predictor.
Comparing the data of Table~\ref{tabsecant}
with Table~\ref{tabquadr}, we observe significant
differences in the average and minimal step sizes along a path.
Timings in Table~\ref{tabsecant} and~\ref{tabquadr} are
for runs on eight cores.

\begin{table}[hbt]
\begin{center}
\begin{tabular}{c||r|r|r|r|r}
\multicolumn{6}{c}{\tt 20-by-20 systems, monomials of degree 10, secant predictor} \\
 & \multicolumn{1}{c|}{\#succ. corrs} &
\multicolumn{1}{c|}{\#corrs} & \multicolumn{1}{c|}{time} &
\multicolumn{1}{c|}{avg step}&  min step \\ \hline
Syst. 1  & {\tt 141244} & {\tt 142657} & {\tt 3h55m3.559s} & 7.28E-06 & 1.22E-06 \\
Syst. 2 & {\tt 176512} & {\tt 178274} & {\tt 8h34m5.082s} & 5.83E-06& 3.05E-07 \\
Syst. 3 & {\tt 150112} & {\tt 151612} & {\tt 7h5m29.649s} & 6.85E-06& 3.05E-07 \\
Syst. 4 & {\tt 125231} & {\tt 126483} & {\tt 7h26m11.352s} & 8.21E-06& 1.22E-06\\
Syst. 5 & {\tt 187772} & {\tt 189645} & {\tt 9h19m29.869s} & 5.48E-06& 3.05E-07 \\
Average & {\tt 156174.2} & {\tt 157734.2} & {\tt 7h16m7.900s} & 6.73E-06& 6.71E-07 \\
St. Dev. & {\tt 25637.81} & {\tt 25892.2} & {\tt 2h4m31.303s} & 1.11E-06& 5.01E-07 \\
\end{tabular}
\caption{For five differently generated systems, we respectively
report the number of successful corrector stages, the total number of 
corrections, the total time, the average and minimal step size along
a solution path, using a secant predictor.}
\label{tabsecant}
\end{center}
\end{table}

\begin{table}[hbt]
\begin{center}
\begin{tabular}{c||r|r|r|r|r}
\multicolumn{6}{c}{\tt 20-by-20 systems, monomials of degree 10, quadratic predictor} \\
 & \multicolumn{1}{c|}{\#succ.corrs} &
\multicolumn{1}{c|}{\#corrs} & \multicolumn{1}{c|}{time} &
\multicolumn{1}{c|}{avg step}&  min step \\ \hline
Syst. 1  & {\tt 571} & {\tt 624} & {\tt 0m59.552s} & 1.91E-03 & 3.13E-04 \\
Syst. 2 & {\tt 791} & {\tt 864} & {\tt 2m34.505s} & 1.37E-03& 7.81E-05 \\
Syst. 3 & {\tt 668} & {\tt 730} & {\tt 2m4.725s} & 1.62E-03& 3.91E-05 \\
Syst. 4 & {\tt 528} & {\tt 578} & {\tt 1m39.336s} & 2.06E-03& 1.56E-04 \\
Syst. 5 & {\tt 848} & {\tt 924} & {\tt 2m39.015s} & 1.27E-03& 7.81E-05 \\
Average & {\tt 681.2} & {\tt 744} & {\tt 1m59.427s} & 1.65E-03& 1.33E-04 \\
St. Dev. & {\tt 137.538} & {\tt 149.124} & {\tt 0m41.275s} & 3.39E-04& 1.09E-04 \\
\end{tabular}
\caption{For five differently generated systems, we respectively
report the number of successful corrector stages, the total number of 
corrections, the total time, the average and minimal step size along
a solution path, using a quadratic predictor.}
\label{tabquadr}
\end{center}
\end{table}

\section{Faster Evaluation of Polynomials and their Derivatives}

In this section, we describe our application of techniques
of algorithmic differentiation~\cite{GW08}.

Along with each normalized monomial 
$x_{i_1}^{a_1} x_{i_2}^{a_2} \cdots x_{i_k}^{a_k}$ with
$1 \leq i_1 < i_2 < \ldots i_k \leq n$ and $a_1,\ldots,a_k \geq 1$,
which appears in the original homotopy $H(\x,t) = \zero$, 
there appear associated to it monomials 
$x_{i_1}^{a_1-1} x_{i_2}^{a_2} \cdots
 x_{i_k}^{a_k}$, $x_{i_1}^{a_1} x_{i_2}^{a_2-1}
\cdots x_{i_k}^{a_k}$, $\ldots$, 
$x_{i_1}^{a_1} x_{i_2}^{a_2} \cdots x_{i_k}^{a_k-1}$ 
in $\frac{\partial H}{\partial {x_{i_1}}}$,
$\frac{\partial H}{\partial x_{i_2}}$, $\ldots$, 
$\frac{\partial H}{\partial x_{i_k}}$ respectively.
Employing the fact that the exponents of the monomial partial 
derivatives do not differ much from the exponents
of the original monomial, we compute the values of the original monomial 
and of the k associated to it monomials in partial derivatives as follows:
\begin{enumerate}
  \item We first compute the common factor
        $x_{i_1}^{a_1-1} x_{i_2}^{a_2-1} \cdots x_{i_k}^{a_k-1}$ 
        of the monomial and its derivatives.
  \item We multiply the value of the common factor by
        $\displaystyle \omega_m=
          \mathop{\prod_{j=1}^{j=k}}_{j \not= m} x_{i_j} \,\,\,$, 
        $m = 1,2,\ldots,k$,
        to get the values of the monomial partial derivatives. 
  \item We multiply 
        $x_{i_1}^{a_1-1} x_{i_2}^{a_2} \cdots x_{i_k}^{a_k}$ by
        $x_{i_1}$ to obtain the original monomial.
\end{enumerate}
The products $\omega_m$, for $m = 1,2,\ldots,k$ we
obtain in $3k-6$ multiplications in the following fashion:
\begin{enumerate}
  \item Recursively we get all products 
        $\psi_m=x_{i_1} x_{i_2} \cdots x_{i_m}, m = 1,2,\ldots,k-1$ 
        by $\psi_m=\psi_{m-1} x_{i_m}$, $\psi_1=x_{i_1}$.
  \item Similarly we obtain all products 
        $\varphi_m=x_{i_k} x_{i_{k-1}} \cdots x_{i_{k-m+1}}$,
        $m = 1,2,\ldots,k-1 $ by
        $\varphi_m=\varphi_{m-1} x_{i_{k-m+1}}$, $\varphi_1=x_{i_k}$.
  \item Finally we get the products $\omega_m \,\,\, m = 1,2,\ldots,k $ as
        $\omega_1=\psi_{k-1}$, $\omega_k=\varphi_{k-1}$, 
        $\omega_m = \psi_{m-1} \varphi_{k-m+2}$, $m = 2, \ldots, k-1$.
\end{enumerate}
In~\cite{GW08}, the evaluation of all derivatives of a product
of variables is known as Speelpenning's example.
Because we assume that our polynomials are sparse,
we may focus on the individual monomials.
For dense polynomials, a nested Horner scheme would be
more appropriate.

\section{Computational Experiments}

The code was developed on a Mac OS X
computer with two 3.2Ghz quad core Intel Xeon processors.
For multithreading, we use the standard {\tt pthreads} library
and QD-2.3.9 for the quad double arithmetic.

In Table~\ref{tabmthomotopy40} we list one generated example for
the complete integrated multithreaded version of the path tracker,
for a system of dimension 40, once with polynomials of degree~2
and once with polynomial of degree~20.
In the latter case, we get a close to optimal speedup
and also for quadratic polynomials, the speedup is acceptable.
Comparing with Table~\ref{tabmthomotopy20}, we see that the degree
of the polynomials are the determining factor in achieving a good
speedup.

\begin{table}[hbt]
\begin{center}
\begin{tabular}{c|rrr|c}
\multicolumn{5}{c}{\tt Dim=40, 40 monomials of degree 2 in a polynomial} \\
\#threads & \multicolumn{1}{c}{real} &
\multicolumn{1}{c}{user} & \multicolumn{1}{c|}{sys} & speedup \\ \hline
1 & {\tt 5m25.509s} & {\tt 5m25.240s} & {\tt 0m0.254s} & 1 \\
2 & {\tt 2m54.098s} & {\tt 5m47.506s} & {\tt 0m0.186s} & 1.870\\
4 & {\tt 1m38.316s} & {\tt 6m31.580s} & {\tt 0m0.206s} & 3.312 \\
8 & {\tt 1m 2.257s} & {\tt 8m11.130s} & {\tt 0m0.352s} & 5.226 \\ \hline
\multicolumn{5}{c}{\tt Dim=40, 40 monomials of degree 20 in a polynomial} \\
\#threads & \multicolumn{1}{c}{real} &
\multicolumn{1}{c}{user} & \multicolumn{1}{c|}{sys} & speedup \\ \hline
1 & {\tt 244m55.691s} & {\tt 244m48.501s} & {\tt 0m 6.621s} & 1 \\
2 & {\tt 123m 1.536s} & {\tt 245m53.987s} & {\tt 0m 3.838s} & 1.991 \\
4 & {\tt  61m53.447s} & {\tt 247m14.921s} & {\tt 0m 4.181s} & 3.958 \\
8 & {\tt  32m22.671s} & {\tt 256m27.142s} & {\tt 0m11.541s} & 7.567 \\
\end{tabular}
\caption{Elapsed real, user, system time, and speedup for tracking
one path in complex quad double arithmetic on a system of dimension~40,
once with quadrics, and once with polynomials of degree~20.}
\label{tabmthomotopy40}
\end{center}
\end{table}

\begin{table}[hbt]
\begin{center}
\begin{tabular}{c|rrr|c}
\multicolumn{5}{c}{\tt Dim=20, 20 monomials of degree 2 in a polynomial} \\
\#threads & \multicolumn{1}{c}{real} &
\multicolumn{1}{c}{user} & \multicolumn{1}{c|}{sys} & speedup \\ \hline
1 & {\tt 0m37.853s} & {\tt 0m37.795s} & {\tt 0m0.037s} & 1 \\
2 & {\tt 0m21.094s} & {\tt 0m42.011s} & {\tt 0m0.063s} & 1.794 \\
4 & {\tt 0m12.804s} & {\tt 0m50.812s} & {\tt 0m0.061s} & 2.956 \\
8 & {\tt 0m 8.721s} & {\tt 1m 8.646s} & {\tt 0m0.097s} & 4.340 \\
\\
\multicolumn{5}{c}{\tt Dim=20, 20 monomials of degree 10 in a polynomial} \\
\#threads & \multicolumn{1}{c}{real} &
\multicolumn{1}{c}{user} & \multicolumn{1}{c|}{sys} & speedup \\ \hline
1 & {\tt 7m17.758s} & {\tt 7m17.617s} & {\tt 0m0.123s} & 1 \\
2 & {\tt 3m42.742s} & {\tt 7m24.813s} & {\tt 0m0.206s} & 1.965\\
4 & {\tt 1m53.972s} & {\tt 7m34.386s} & {\tt 0m0.150s} & 3.841 \\
8 & {\tt 0m59.742s} & {\tt 7m53.469s} & {\tt 0m0.279s} & 7.327 \\ \hline
\end{tabular}
\end{center}
\caption{Elapsed real, user, system time, and speedup for tracking
one path in complex quad double arithmetic on a system of dimension~20,
once with quadrics, and once with polynomials of degree~10.}
\label{tabmthomotopy20}
\end{table}

The results in Tables~\ref{tabmthomotopy40}
and~\ref{tabmthomotopy20} are done without the faster
evaluation of polynomials and their derivatives,
so the degrees matter most in the speedup.
With faster evaluation routines, the threshold on the dimension
for the speedup will have to be higher.

We end this paper with some preliminary sequential timings
on using faster evaluation and differentiation schemes.
In our previous implementation, the continuation parameter~$t$
was treated as just another variable and this led to an overhead
of a factor~3.  Table~\ref{tabnew_diff} 
contains experimental results.

\begin{table}[hbt]
\begin{center}
\begin{tabular}{c||r|r|r|c}
\multicolumn{5}{c}{\tt 20-by-20 systems, polynomial evaluation, 400 times, sequential execution} \\
degrees & \multicolumn{1}{c|}{new alg.time} &
\multicolumn{1}{c|}{old alg.time} & \multicolumn{1}{c|}{speedup} &
\multicolumn{1}{c}{pure speedup} \\ \hline
 5  & {\tt 4.267s} & {\tt 36.521s} & {\tt 8.59} & 2.85 \\
 10 & {\tt 7.280s} & {\tt 1m7.114s} & {\tt 9.22} & 3.07 \\
 20 & {\tt 11.304s} & {\tt 2m3.122s} & {\tt 10.89} & 3.63  \\
\multicolumn{5}{c}{\tt 40-by-40 systems, polynomial evaluation, 400 times, sequential execution} \\
degrees & \multicolumn{1}{c|}{new alg.time} &
\multicolumn{1}{c|}{old alg.time} & \multicolumn{1}{c|}{speedup} &
\multicolumn{1}{c}{pure speedup} \\ \hline
 5  & {\tt 19.855s} & {\tt 4m58.162s} & {\tt 15.02} & 5.01 \\
 10 & {\tt 36.737s} & {\tt 8m48.765s} & {\tt 14.39} & 4.80 \\
 20 & {\tt 1m1.980ss} & {\tt 16m4.541s} & {\tt 15.56} & 5.19  \\
\end{tabular}
\caption{For a system of dimension~20 and~40,
for increasing degrees, we list the times of the new algorithm,
the old algorithm and the speedup.  The pure speedup is the speedup
divided by 3, to account for treating the continuation parameter~$t$
differently.}
\label{tabnew_diff}
\end{center}
\end{table}

\section{Conclusions}

For polynomial systems where the computational cost is dominated by
the evaluation of polynomials, the multithreaded version of our
path tracker performs already well in modest dimensions.
For systems of lower degrees, the threshold dimension for good speedup
will need to be higher because then the cost of Gaussian elimination
becomes more important.

\bibliographystyle{plain}

\begin{thebibliography}{10}

\bibitem{Akl04}
S.G. Akl.
\newblock Superlinear performance in real-time parallel computation.
\newblock {\em The Journal of Supercomputing}, 29(1):89--111, 2004.

\bibitem{DKK03}
Y.~Dai, S.~Kim, and M.~Kojima.
\newblock Computing all nonsingular solutions of cyclic-n polynomial using
  polyhedral homotopy continuation methods.
\newblock {\em J. Comput. Appl. Math.}, 152(1-2):83--97, 2003.

\bibitem{Dek71}
T.J. Dekker.
\newblock A floating-point technique for extending the available precision.
\newblock {\em Numerische Mathematik}, 18(3):224--242, 1971.

\bibitem{GW08}
A.~Griewank and A.~Walther.
\newblock {\em Evaluating Derivatives: Principles and Techniques of Algorithmic
  Differentiation}.
\newblock SIAM, second edition, 2008.

\bibitem{HLB00}
Y.~Hida, X.S. Li, and D.H. Bailey.
\newblock Algorithms for quad-double precision floating point arithmetic.
\newblock In {\em {15th IEEE Symposium on Computer Arithmetic (Arith-15 2001),
  11-17 June 2001, Vail, CO, USA}}, pages 155--162. IEEE Computer Society,
  2001.
\newblock Shortened version of Technical Report LBNL-46996, software at {\tt
  http://crd.lbl.gov/$\sim$dhbailey/} {\tt mpdist/qd-2.3.9.tar.gz}.

\bibitem{LLT08}
T.L. Lee, T.Y. Li, and C.H. Tsai.
\newblock {HOM4PS-2.0}: a software package for solving polynomial systems by
  the polyhedral homotopy continuation method.
\newblock {\em Computing}, 83(2-3):109--133, 2008.

\bibitem{LVZ06}
A.~Leykin, J.~Verschelde, and A.~Zhao.
\newblock Newton's method with deflation for isolated singularities of
  polynomial systems.
\newblock {\em Theoretical Computer Science}, 359(1-3):111--122, 2006.

\bibitem{Rum10}
S.M. Rump.
\newblock Verification methods: Rigorous results using floating-point
  arithmetic.
\newblock {\em Acta Numerica}, 19:287–449, 2010.

\bibitem{Ver99}
J.~Verschelde.
\newblock Algorithm 795: {PHC}pack: A general-purpose solver for polynomial
  systems by homotopy continuation.
\newblock {\em ACM Trans. Math. Softw.}, 25(2):251--276, 1999.
\newblock Software available at {\tt
  http://www.math.uic.edu/{\~{}}jan/download.html}.

\bibitem{VY10}
J.~Verschelde and G.~Yoffe.
\newblock Polynomial homotopies on multicore workstations.
\newblock In M.M. Maza and J.-L. Roch, editors, {\em Proceedings of the 2010
  International Workshop on Parallel Symbolic Computation (PASCO 2010), July
  21-23 2010, Grenoble, France}, pages 131--140. ACM, 2010.

\end{thebibliography}

\end{document}